\documentclass[twocolumn, nofootinbib, nobibnotes, amsmath,amssymb,aps, pra, floatfix]{revtex4-1}

\usepackage{bbm}
\usepackage{mathtools}
\usepackage[export]{adjustbox}
\usepackage{wrapfig}
\usepackage{import}
\usepackage{bbold}
\usepackage{microtype}
\usepackage{bm} 
\usepackage{graphicx}% Include figure files
\usepackage{dcolumn}% Align table columns on decimal point
\usepackage{color}
\usepackage{silence}
\renewcommand{\v}[1]{\ensuremath{\mathbf{#1}}} % for vectors
\newcommand{\gv}[1]{\ensuremath{\mbox{\boldmath$ #1 $}}} % for vectors of Greek letters
\newcommand{\pd}[2]{\frac{\partial #1}{\partial #2}} % for partial derivatives
\newcommand{\grad}[1]{\gv{\nabla} #1} % for gradient
\renewcommand{\div}[1]{\gv{\nabla} \cdot #1} % for divergence
\begin{document}

\preprint{APS/123-QED}

\title{Gauge transformations and Galilean covariance in nonlinear gauge-coupled quantum fluids}% Force line breaks with \\

\author{Yvan Buggy}
%\altaffiliation[CM-CDT]{}
%\altaffiliation[CM-CDT]{}
 \author{Patrik {\"O}hberg}
 %\email{P.Ohberg@hw.ac.uk}
\affiliation{SUPA, Institute of Photonics and Quantum Sciences, Heriot-Watt University, Edinburgh EH14 4AS, United Kingdom}

%\date{\today}

\begin{abstract}
    We investigate certain invariance properties of quantum fluids subject to a nonlinear gauge potential.
	In particular, we derive the covariant transformation laws for the nonlinear potentials under a space-time Galilean boost and consider $U\left(1\right)$ gauge transformations.
    We find that the hydrodynamic canonical field equations are form-invariant in the case of external gauge functions $\chi(\v{r}, t)$, but not for nonlinear gauge functionals $\chi[\rho]$.
    Hence, nonlinear gauge potentials are non-trivial potentials which may not be "gauged-away".
    Notably, for a superfluid in dimension $d=1$, attempting to do so generates the gauge-pressure of the fluid in the Hamiltonian density.
    Further, we investigate how the field equations transform under arbitrary Galilean transformations.
    We find that the immediate lack of Galilean covariance is restored under a suitably chosen transformation rule set for the potentials, which is identical in form to that of a Schr\"odinger particle coupled to external scalar and vector potentials.

%\begin{description}
%\end{description}
\end{abstract}

\maketitle

%\tableofcontents
\section{Introduction}
The pursuit for implementing artificial electromagnetic potentials for charge neutral systems has received a steady stream of interest over the last two decades.
These potentials are generally engineered through combined-interactions, such that a system exhibits spatially varying local eigenstates, which, in turn, impart a geometric phase onto the wavefunction \cite{dalibard2011colloquium,goldman2014light,berry1984quantal,peskin1989aharonov}.
Although initial implementations involved static synthetic fields, further proposals have since been put forward for creating "dynamical" gauge fields which are described by their own Hamiltonian and not merely imposed on the system \cite{kapit2011optical,wiese2013ultracold,banerjee2012atomic,tagliacozzo2013simulation,zohar2013cold,tagliacozzo2013optical}.
One possibility for realising a dynamical gauge potential, albeit not a field in the strict sense, is to introduce a "back-action" where the dynamics of the gauge potential are tied to the motion of the condensate.
Such a prospect has been proposed in \cite{edmonds2013,edmonds2013josephson,keilmann2011statistically,greschner2014density,cardarelli2016engineering,PhysRevLett.115.053002,PhysRevLett.117.205303}, where the effective field depends on the spatial configuration of the atoms, in particular, the atomic density.
More recently, a density-dependent gauge potential has been experimentally demonstrated in a two-dimensional optical lattice, by modulating the interaction strength in synchrony with the lattice shaking \cite{PhysRevLett.121.030402}, where the tunnelling rate depends on the occupation number.

Density-dependent gauge potentials, which we shall refer to as nonlinear gauge potentials, have been shown to give rise to a number of interesting properties, such as chiral solitons \cite{edmonds2013,dingwall2018non,dingwall2019stability} and anyonic structures \cite{aglietti1996anyons}. The superfluid properties can also be dramatically affected by the gauge potential, where for instance drag forces will become directionally dependent \cite{buggy2019}. From a dynamical perspective, perhaps the most striking feature of a density-dependent gauge potential is the occurrence of an exotic flow-dependent nonlinear term in the wave equation for the quantum phase.
This is exemplified notably by the stress tensor of the fluid, which features a "gauge pressure" term related to the overlap of the gauge potential and the gauge-covariant current \cite{buggy2020hydrodynamics}.
One would expect such a term to carry important implications for the symmetry properties of the fluid.
For instance, a flow-dependent fluid pressure evidently transforms between Galilean frames.
This suggests that additional transformation rules should be implemented in order to "restore" the covariance of the dynamical equations under a Galilean boost.

In this paper, we derive these transformation rules and also investigate the invariance of the dynamical equations of such systems under both external and nonlinear gauge transformations.
We work within a hydrodynamic canonical formalism and consider transformations at the level of the field equations, rather than the wave equations proceeding from these.
Furthermore, although the motivation for our study stems from a well-established microscopic model proposed in \cite{edmonds2013}, we follow a general approach adopted from a previous paper \cite{buggy2020hydrodynamics}, which is not tied to any particular model.
Here, we simply assume that a density-dependent gauge potential $\v{A}\left(\rho\right)$ enters the Hamiltonian functional, whose components take the form of arbitrary density-dependent functions.
Hence, our concern lies not with the physical origin of the potentials, but rather, the formal structure of the dynamical equations which result from such a coupling.
We shall refer to this class of fluids as \textit{nonlinear gauge-coupled quantum fluids}. 

The paper is outlined as follows.
In section \ref{sec:hydro}, we begin by reviewing the canonical field formalism and hydrodynamic equations for such fluids, which will be used subsequently in the text.
Further, in section \ref{sec:gauge}, we consider gauge transformations, treating both the case of external gauge functions and nonlinear gauge functionals.
Finally, in section \ref{sec:galilean}, we investigate the covariance of the canonical field equations under general Galilean transformations.

\section{\label{sec:hydro}Hydrodynamic equations of the fluid}
Our study pertains to the class of nonlinear quantum fluids whose effective meanfield Hamiltonian may be written in the form
\begin{equation}
	\hat{H}_{MF} = \frac{\left( \hat{\v{p}}-\v{A}\left( \rho \right) \right)^2}{2m}+\eta\left( \rho \right),
	\label{eq:meanfieldHamiltonianNonlinearGaugeCoupledQuantumFluid}
\end{equation}
where $\eta$ and $\v{A}$ are nonlinear effective potentials which depend on the density of the fluid.
After performing a Madelung transformation, $\psi=\sqrt{\rho}e^{i\theta/\hbar}$, on the macroscopic condensate wavefunction $\psi$, it is possible to show \cite{buggy2020hydrodynamics} that the dynamics of the field components $\rho$ and $\theta$, may be expressed in the Hamiltonian form 
\begin{align}
  \dot{\rho}\left( \v{r} \right)&-\frac{\delta H}{\delta\theta\left( \v{r} \right)}=0, \label{eq:canonicalFieldEquationRhoDotSymmetry} \\
  \dot{\theta}\left( \v{r} \right)&+\frac{\delta H}{\delta\rho\left( \v{r} \right)}=0, \label{eq:canonicalFieldEquationThetaDotSymmetry}
\end{align}
where the Hamiltonian functional of the field, is given by 
\begin{equation}
	H\left[ \rho,\theta \right]=\int d^3\v{r}\left\{  \rho\left[\frac{\left( \grad{\theta}-\v{A}\left( \rho \right) \right)^2}{2m}+\eta\left( \rho \right)\right]+\mathcal{Q}\right\},
	\label{eq:HamiltonianDensityNonlinearGaugeCoupledQuantumFluid}
\end{equation}
with $\mathcal{Q}=\left(\hbar\grad{\rho}\right)^2/\left( 8m\rho \right)$ representing a quantum energy density.
Note that the derivatives of the Hamiltonian appearing in Eqs. (\ref{eq:canonicalFieldEquationRhoDotSymmetry}) and (\ref{eq:canonicalFieldEquationThetaDotSymmetry}) are functional, or variational derivatives.
As mentioned previously, a defining feature of a density-dependent gauge potential is the occurrence of a nonlinear flow term in the wave equation for $\theta$.
Indeed, substituting the above expression for $H$ into the canonical field equations (\ref{eq:canonicalFieldEquationRhoDotSymmetry}) and (\ref{eq:canonicalFieldEquationThetaDotSymmetry}), yields, respectively, the wave equations
\begin{align}
	\partial_t\rho&+\div{\left(\rho\v{v}\right)}=0, \label{eq:waveEquationRhoNLGCQF} \\
	\partial_t\theta&+\frac{1}{2}mv^2+\Phi\left( \rho,\v{u} \right)+Q=0, \label{eq:waveEquationThetaNLGCQF}
\end{align}
where 
\begin{equation}
    \v{v}=\v{u}-\v{A}/m,
    \label{eq:flow_covariant}
\end{equation}
is the physical velocity or mechanical flow, $\v{u}=\grad{\theta}/m$ is the canonical or phase flow and $Q=-\hbar^2\nabla^2\sqrt{\rho}/\left( 2m\sqrt{\rho} \right)$ is the quantum potential.
Moreover, the effective potentials lead to a non-trivial nonlinear term 
\begin{equation}
	\Phi=\eta+\rho\pd{\eta}{\rho}-\rho\v{v}\cdot\pd{\v{A}}{\rho},
	\label{eq:nonlinearScalarPotentialWaveEquationTheta}
\end{equation}
in the wave equation of the phase, which is linear in the canonical flow $\v{u}$.
As a consequence, the fluid pressure also inherits a flow-dependent term.
This may be seen by expressing the hydrodynamic equations (\ref{eq:waveEquationRhoNLGCQF}) and (\ref{eq:waveEquationThetaNLGCQF}) in the reference frame of the fluid, which yield \cite{buggy2020hydrodynamics}, respectively
\begin{align}
  &\left( \pd{}{t}+\v{v}\cdot\grad{} \right)\rho+\rho\div{\v{v}}=0, \label{eq:waveEquationRhoSecondForm}\\
&m\rho\left( \pd{}{t}+\v{v}\cdot\grad{} \right)u_k=\nabla_j\left( -\delta_{jk}P+\sigma_{jk} \right), \label{eq:cauchyEquation}
\end{align}
where  
\begin{equation}
  \sigma_{jk}=-\frac{\hbar^2}{4m\rho}\nabla_j\rho\nabla_k\rho,
  \label{eq:quantumStressTensor}
\end{equation}
is the quantum stress tensor, and $P$ is the fluid pressure, given by 
\begin{equation}
    P = -\frac{\hbar^2}{4m}\nabla^2\rho+\rho\left(\Phi-\eta\right).
    \label{eq:pressureNLGCQF}
\end{equation}
Note the distinction between $\v{v}$ and $\v{u}$ in equation (\ref{eq:cauchyEquation}).
As such, this equation describes the transport of \textit{canonical} momentum rather than mechanical momentum.
Recalling expression (\ref{eq:nonlinearScalarPotentialWaveEquationTheta}) for $\Phi$, we see that $P$ depends explicitly on the flow profile of the fluid.
In particular, the fluid pressure may be viewed as a combination of the following three terms:
\begin{equation}
    P = P_Q + P_{\eta} + P_{\v{A}}, 
    \label{eq:pressureNLGCQF_combination}
\end{equation}
where, $P_Q=-\hbar^2\nabla^2\rho/(4m)$, is the quantum pressure, $P_{\eta} = \rho^2\partial_{\rho}\eta$, is the pressure contribution from the nonlinear scalar potential, and, $P_{\v{A}} = -\rho^2\v{v}\cdot\partial_{\rho}\v{A}$, is the pressure contribution from the nonlinear gauge potential, which we refer to as the \textit{gauge pressure}.

\section{\label{sec:gauge}Gauge transformations}
Let us investigate the implications of a nonlinear gauge potential on the symmetry properties of the fluid.
To begin with, we consider gauge transformations of the form
\begin{align}
	\theta\rightarrow&\theta'=\theta+\chi, \label{eq:phaseTransformationGaugeTransformation} \\
	\v{A}\rightarrow&\v{A}'=\v{A}+\grad{\chi}, \label{eq:gaugePotentialTransformationGaugeTransformation}\\
    \eta\rightarrow&\eta'=\eta-\partial_t\chi. \label{eq:scalarPotentialTransformationGaugeTransformation}
\end{align}
We shall treat both the case in which $\chi$ is an external single-valued scalar function of space and time, $\chi=\chi\left(\v{r}, t\right)$, and a functional of the density, $\chi=\chi\left[\rho\right]$. 
Notice that the latter is required for evaluating whether nonlinear gauge potentials are physical potentials which may not be trivially gauged away, since any attempt to do so involves a gauge functional given by the contour integral of some function of the density. 
In both cases, the transformed canonical field equations may be written in the generic form 
\begin{align}
  \dot{\rho}\left( \v{r} \right)&-\frac{\delta H'}{\delta\theta'\left( \v{r} \right)}=0, \label{eq:canonicalFieldEquationRhoDotGaugeTransformed}\\
  \dot{\theta}'\left( \v{r} \right)&+\frac{\delta H'}{\delta\rho\left( \v{r} \right)}=0, \label{eq:canonicalFieldEquationThetaDotGaugeTransformed}
\end{align}
where the Hamiltonian transforms according to 
\begin{equation}
    H'=\int d^3\v{r}\left\{ \rho\left[ \frac{\left( \grad{\theta}'-\v{A}' \right)^2}{2m}+\eta' \right]+\mathcal{Q} \right\}.
  \label{eq:HamiltonianGaugeTransformedSchrodingerField}
\end{equation}
In the Madelung representation of the fluid, it becomes readily apparent that the combination of transformations (\ref{eq:phaseTransformationGaugeTransformation}) and (\ref{eq:gaugePotentialTransformationGaugeTransformation}) leaves the mechanical flow $\v{v}$ from Eq. (\ref{eq:flow_covariant}) unchanged, i.e. $\v{v}$ is gauge invariant.
Hence, according to Eq. (\ref{eq:HamiltonianGaugeTransformedSchrodingerField}), the new and old Hamiltonians are related by
\begin{equation}
  H'\left[ \rho,\theta' \right]=H\left[ \rho,\theta \right]-\int d^3\v{r}\rho\pd{\chi}{t}.
  \label{eq:HamiltonianRelationshipGaugeTransformation}
\end{equation}

\subsection{\label{sec:externalGaugeFunctions}External gauge functions $\chi\left(\v{r}, t\right)$}
Let us examine the situation where $\chi$ is an externally prescribed scalar function.
Clearly, in this case, $\delta\chi=0$ for arbitrary field variations $\delta\rho$ and $\delta\theta$.
Hence, the first variation of Eq. (\ref{eq:HamiltonianRelationshipGaugeTransformation}), yields
\begin{align}
  \frac{\delta H'}{\delta\theta'\left( \v{r} \right)}&= \frac{\delta H}{\delta\theta\left( \v{r} \right)}, \label{eq:variationalDerivativeHamiltonianRelationTheta}\\
  \frac{\delta H'}{\delta\rho\left( \v{r} \right)}&= \frac{\delta H}{\delta\rho\left( \v{r} \right)}-\pd{\chi\left( \v{r},t \right)}{t}, \label{eq:variationalDerivativeHamiltonianRelationRho}
\end{align}
Inserting the first and second of these into, respectively, the transformed field Eqs. (\ref{eq:canonicalFieldEquationRhoDotGaugeTransformed}) and (\ref{eq:canonicalFieldEquationThetaDotGaugeTransformed}), we immediately recover the original field Eqs. (\ref{eq:canonicalFieldEquationRhoDotSymmetry}) and (\ref{eq:canonicalFieldEquationThetaDotSymmetry}).
Therefore, the canonical equations of the nonlinear fluid are form invariant under gauge transformations generated by external gauge functions $\chi\left(\v{r},t\right)$.
Inspecting the form of the nonlinear term in the wave equation for the phase, $\Phi$ from Eq. (\ref{eq:nonlinearScalarPotentialWaveEquationTheta}), this is not surprising.
Indeed, the nonlinear flow term results from a particular form of density-dependence of $\v{A}$, which remains unchanged by an external gauge function.

\subsection{\label{sec:nonlinearGaugeFunctionals}Nonlinear gauge functionals $\chi\left[\rho\right]$}
Next we examine the case where $\chi$ takes the form of some functional of the density.
To begin with, observe how the canonical structure of the field equations (\ref{eq:canonicalFieldEquationRhoDotSymmetry}) and (\ref{eq:canonicalFieldEquationThetaDotSymmetry}),  implies that the field components $\rho$ and $\theta$ play the role of conjugate variables, where the Poisson bracket of two dynamical variables $f$ and $g$ on phase space, reads
\begin{multline}
    \left\{ f\left( \v{x} \right),g\left( \v{y} \right) \right\}_{\rho,\theta}\\
    =\int d^3\v{r}\left( \frac{\delta f\left( \v{x} \right)}{\delta\rho\left( \v{r} \right)}\frac{\delta g\left( \v{y} \right)}{\delta\theta\left( \v{r} \right)}-\frac{\delta f\left( \v{x} \right)}{\delta\theta\left( \v{r} \right)}\frac{\delta g\left( \v{y} \right)}{\delta\rho\left( \v{r} \right)} \right).
  \label{eq:poissonBracketReducedPhaseSpace}
\end{multline}
Now, for gauge transformations generated by $\chi\left[\rho\right]$, the field variables transform according to
\begin{equation}
  \begin{pmatrix}
    \rho \\ \theta
  \end{pmatrix}
  \rightarrow
  \begin{pmatrix}
    \rho'=\rho \\ \theta'=\theta+\chi\left[ \rho \right]
  \end{pmatrix}.
  \label{}
\end{equation}
Let us check whether the new variables form a canonically conjugate pair.
To this end, we evaluate the Poisson bracket of the new variables with respect to the old variables, which gives
\begin{multline}
  \left\{ \rho'\left( \v{x}\right), \theta'\left( \v{y}  \right) \right\}_{\rho,\theta}\\
  =\int d^3\v{r}\left( \frac{\delta\rho'\left( \v{x} \right)}{\delta\rho\left( \v{r} \right)}\frac{\delta\theta'\left( \v{y} \right)}{\delta\theta\left( \v{r} \right)}-\frac{\delta\rho'\left( \v{x} \right)}{\delta\theta\left( \v{r} \right)}\frac{\delta\theta'\left( \v{y} \right)}{\delta\rho\left( \v{r} \right)} \right)\\
  =\delta\left( \v{x}-\v{y} \right),
  \label{}
\end{multline}
on account of $\rho'$ being independent of $\theta$.
Hence $\theta'$ and $\rho$ may be treated as independent variables.
We shall make use of this property further in section \ref{sec:oneDimSuperfluid}.

Following the same line of reasoning as section \ref{sec:externalGaugeFunctions}, let us evaluate the first variation of Eq. (\ref{eq:HamiltonianRelationshipGaugeTransformation}) for the case of a nonlinear gauge functional.
Since $\chi$ is independent of $\theta$, Eq. (\ref{eq:variationalDerivativeHamiltonianRelationTheta}) remains valid and the transformed canonical equation (\ref{eq:canonicalFieldEquationRhoDotGaugeTransformed}) for $\rho$ again reduces to the original canonical equation (\ref{eq:canonicalFieldEquationRhoDotSymmetry}).
However, the canonical equation for $\theta$ is no longer form invariant, but transforms according to
\begin{equation}
  \dot{\theta}'+\frac{\delta H}{\delta\rho}-\frac{\delta}{\delta\rho}\int d^3\v{r}\rho\partial_t\chi\left[ \rho \right]=0,
  \label{eq:canonicalFieldEquationThetaDotTransformedGaugeTransformedTwo}
\end{equation}
where an additional term appears due to $\chi\left[ \rho \right]$.
Although the form taken by this term will depend on $\chi\left[ \rho \right]$, one may see that current terms will generally be involved, through $\partial_t\chi\left[ \rho \right]$.
In the following section, we examine one such gauge functional.
We may therefore conclude that density-dependent gauge potentials are physical potentials which may not be trivially gauged away.
\subsection{\label{sec:oneDimSuperfluid}The one-dimensional gauge-coupled superfluid}
Typically, one may be interested in gauge functionals which eliminate the gauge potential from the kinetic energy density.
Let us examine one such case.
As our system, we consider the superfluid fraction of particles in an optically-addressed weakly-interacting dilute Bose gas of two-level atoms, first proposed in \cite{edmonds2013}.
Confining the dynamics to one of the condensate components, a density-modulated gauge potential emerges in the form $\v{A}=\v{a}\rho$, where $\rho$ is the atomic density of the relevant condensate component, while $\v{a}=\left(g_{11}-g_{22}\right)\grad{\phi}/\left(8\Omega\right)$ controls the effective strength and orientation of the nonlinear vector potential.
Here, $\Omega$ is the generalised Rabi-frequency of the light-matter coupling, $\phi$ is the phase of the incident laser field and $g_{ij}=4\pi\hbar^2a_{ij}/m$ denote the coupling constants of the channels stemming from pairwise interactions between the atoms, where $a_{ij}$ are the associated s-wave scattering lengths.

The system is described by a wave equation of the form (\ref{eq:waveEquationThetaNLGCQF}), with $\Phi=-\v{a}\cdot\v{J}+g\rho$ and $g=\left(g_{11}+g_{22}+2g_{12}\right)/4$.
For a superfluid confined to dimension $d=1$, the wave equation for the phase reads  
\begin{equation}
  \partial_t\theta+\frac{1}{2}mv^2-aJ+g\rho+Q=0,
  \label{eq:QHJEDensityModulatedGaugePotentialOneDimension}
\end{equation}
where, $v=\left( \partial_x\theta-a\rho \right)/m$ and $J=\rho v$.
We choose $d=1$, since the last term in Eq. (\ref{eq:canonicalFieldEquationThetaDotTransformedGaugeTransformedTwo}) becomes integrable in this case.
In turn, the fluid pressure again takes the form of equation (\ref{eq:pressureNLGCQF_combination}), where the quantum pressure is $P_Q = -\hbar^2\partial_x^2\rho/4m$, $P_{\eta} = g\rho^2/2$, and the gauge pressure reads $P_A=-JA$.

The gauge functional which eliminates $\v{A}$ from the kinetic energy density, is
\begin{equation}
  \chi=-\int_{-\infty}^x dy a \rho\left( y,t \right).
  \label{eq:gaugeFunctionOneDimensionalSystem}
\end{equation}
Invoking this relation in Eq. (\ref{eq:HamiltonianRelationshipGaugeTransformation}), we find that the new and old Hamiltonians are related by
\begin{equation}
  H'=H-\int dx\rho^2 a v',
  \label{eq:HamiltonianGaugeTransformedOneDimensionalSystem}
\end{equation}
where $v'=\partial_x\theta'/m$.
Hence, an additional term $-\int dx J' A=-\int dx J A$, appears in the Hamiltonian under the nonlinear gauge transformation, where $J'=\rho v'$.
In other words, the gauge-pressure, $P_A$, is generated in the Hamiltonian density when one attempts to eliminate the nonlinear gauge potential.
Inserting expression (\ref{eq:HamiltonianGaugeTransformedOneDimensionalSystem}) for $H'$ into the field Eq. (\ref{eq:canonicalFieldEquationThetaDotGaugeTransformed}), yields the wave equation
\begin{equation}
  \partial_t\theta'+\frac{1}{2}m v'v'-2a J'+g\rho+Q=0,
  \label{eq:QHJEDensityModulatedGaugePotentialOneDimensionGaugeTransformed}
\end{equation}
where  we have used $\delta v'/\delta\rho=0$, which holds on account of $\rho$ and $\theta'$ being independent (see section \ref{sec:nonlinearGaugeFunctionals}).
Comparing Eqs. (\ref{eq:QHJEDensityModulatedGaugePotentialOneDimension}) and (\ref{eq:QHJEDensityModulatedGaugePotentialOneDimensionGaugeTransformed}), we see that the wave equation has transformed under the nonlinear gauge transformation.
In particular, the attempt to absorb the nonlinear gauge potential into the phase has produced an additional current term in the wave equation.

\section{\label{sec:galilean}Galilean Transformations}
We now turn our attention to a space-time symmetry group associated with coordinate transformations, namely, Galilean transformations.
The coordinate transformations from some inertial frame of reference $\Sigma$, to another frame $\Sigma'$ moving with a uniform, nonrelativistic relative velocity $\v{w}$, may be written in the general form \cite{holland1995quantum,arnold2007mathematical,iro2015modern}
\begin{align}
  x_i&\rightarrow x_i'=R_{ij}x_j-w_it+a_i, \nonumber\\
  t&\rightarrow t'=t+t_0, \label{eq:Galilean_group}
\end{align}
where $a_i$ and $t_0$ are constants representing offsets in position and time respectively, and $R_{ij}$ is a unitary rotational matrix satisfying $R_{ij}R_{jk}=\delta_{ik}$.
Differential operators also transform between frames under (\ref{eq:Galilean_group}), such that 
\begin{align}
  \nabla_i&\rightarrow \nabla_i'=R_{ij}\nabla_j, \label{eq:differential_operator_transformation_galilean_space} \\
  \pd{}{t}&\rightarrow \pd{}{t'}=\pd{}{t}+w_iR_{ij}\nabla_j. \label{eq:differential_operator_transformation_galilean_time}
\end{align}
In turn, the velocity field, transforms as
\begin{equation}
  v_i\rightarrow v_i'=R_{ij}v_j-w_i.
  \label{eq:velocity_transformation_galilean}
\end{equation}
Since the energy and momentum of the fluid transforms under a Galilean boost, the phase also transforms between reference frames.
Let us denote this transformation, by
\begin{equation}
    \theta\rightarrow\theta'=\theta+\chi,
  \label{eq:phaseTransformationGalilean}
\end{equation}
where $\chi=\chi\left(\v{r},t\right)$ is a single-valued scalar function of space and time.
Furthermore, we should not exclude the prospect that the potentials may be subject to transformations in order that the dynamical equations be form invariant.
In fact, this is already the situation for a nonrelativistic quantum mechanical particle coupled to both external scalar and vector potentials \cite{takagi1991quantum,dewitt1957dynamical,brown1999galilean}. 
Accordingly, let us denote the transformations of the nonlinear potentials, as 
\begin{align}
    \eta&\rightarrow\eta'=\eta+\mu, \label{eq:scalarPotentialTransformationNonlinearGalileanTransformation} \\
    A_i&\rightarrow A_i'=R_{ij}A_j+G_i, \label{eq:vectorPotentialTransformationNonlinearGalileanTransformation}
\end{align}
where $\mu$ and $\v{G}$ are unspecified.
However, in view of the nonlinear character of $\eta$ and $\v{A}$, we allow for the possibility that these also depend on the density.

In the frame $\Sigma'$, the canonical field equations (\ref{eq:canonicalFieldEquationRhoDotSymmetry}) and (\ref{eq:canonicalFieldEquationThetaDotSymmetry}), read
\begin{align}
  \pd{\rho'}{t'}&-\frac{\delta H'}{\delta\theta'}=0, \label{eq:canonicalFieldEquationRhoDotGalileanTransformed} \\
  \pd{\theta'}{t'}&+\frac{\delta H'}{\delta\rho'}=0, \label{eq:canonicalFieldEquationThetaDotGalileanTransformed}
\end{align}
where the Hamiltonian from Eq. (\ref{eq:HamiltonianDensityNonlinearGaugeCoupledQuantumFluid}), transforms to
\begin{equation}
  H'=\int d^3\v{r}'\left[ \rho'\left( \frac{1}{2}m\v{v}'\cdot\v{v}'+\eta'\right)+\mathcal{Q}' \right],
  \label{eq:HamiltonianGalileanTransformedExternalGeneric}
\end{equation}
with $\v{v}'=\left(\grad'\theta'-\v{A}'\right)/m$. 
Since $\rho'=\rho$ and $\v{v}'$ is restricted by Eq. (\ref{eq:velocity_transformation_galilean}), the Hamiltonians in the two frames are related, by
\begin{equation}
    H'\left[ \rho',\theta' \right]=H\left[ \rho,\theta \right]+\int d^3\v{r}\rho\left( \frac{1}{2}mw^2-mw_iR_{ij}v_j+\mu \right).
  \label{eq:HamiltonianGalileanTransformedExternal}
\end{equation}
The first variation of the above equation, yields
\begin{eqnarray}
    \frac{\delta H'}{\delta\rho'}&=& \frac{\delta H}{\delta\rho}+\frac{1}{2}mw^2-mw_iR_{ij}v_j\nonumber\\&&+\mu+\rho\left( \pd{\mu}{\rho}+w_iR_{ij}\pd{A_j}{\rho} \right), \label{eq:variationalDerivativeHamiltonianRelationRhoGalileanNonlinearField}\\
%\end{multline}
%\begin{equation}
  \frac{\delta H'}{\delta\theta'}&=& \frac{\delta H}{\delta\theta}+w_iR_{ij}\nabla_j\rho.\label{eq:variationalDerivativeHamiltonianRelationThetaGalileanNonlinearField}
\end{eqnarray}
Note that in order to obtain Eq. (\ref{eq:variationalDerivativeHamiltonianRelationThetaGalileanNonlinearField}), we have substituted Eq. (\ref{eq:flow_covariant}) for $\v{v}$ into Eq. (\ref{eq:HamiltonianGalileanTransformedExternal}) and integrated by parts accordingly. 
Let us investigate the implications of these relations on the transformed canonical equations (\ref{eq:canonicalFieldEquationRhoDotGalileanTransformed}) and (\ref{eq:canonicalFieldEquationThetaDotGalileanTransformed}).
Invoking Eqs. (\ref{eq:differential_operator_transformation_galilean_time}) and (\ref{eq:variationalDerivativeHamiltonianRelationThetaGalileanNonlinearField}) in Eq. (\ref{eq:canonicalFieldEquationRhoDotGalileanTransformed}), we immediately recover the original field equation for $\rho$.
Hence the canonical equation (\ref{eq:canonicalFieldEquationRhoDotSymmetry}) is form invariant under a Galilean transformation.
Notice here that no demands are being made of $\mu$, $\v{G}$ or $\chi$, other than the restriction that $\chi$ be independent of $\theta$, since the violation of the latter introduces additional terms in Eq. (\ref{eq:variationalDerivativeHamiltonianRelationThetaGalileanNonlinearField}).
Turning our attention to the field equation for $\theta$, let us substitute Eqs. (\ref{eq:differential_operator_transformation_galilean_time}), (\ref{eq:phaseTransformationGalilean}) and (\ref{eq:HamiltonianGalileanTransformedExternal}) into Eq. (\ref{eq:canonicalFieldEquationThetaDotGalileanTransformed}), which gives
\begin{equation}
  \dot{\theta}+\frac{\delta H}{\delta\rho}+\xi=0,
  \label{eq:QHJE_elementary_fluid_galilean_transformed_generic_form}
\end{equation}
where we have grouped the additional terms generated by the Galilean transformation, in the function $\xi=\xi_{\chi}+\xi_{\pi}$, with
\begin{align}
    \xi_{\chi}&= \left(\pd{}{t}+w_iR_{ij}\nabla_j\right)\chi+w_iR_{ij}\left( \nabla_j\theta-mv_j \right)+\frac{1}{2}mw^2,\label{eq:xi_phase_factor}\\
    \xi_{\pi}&=\mu+w_iR_{ij}A_j+\rho\left( \pd{\mu}{\rho}+w_iR_{ij}\pd{A_j}{\rho} \right) \label{eq:xi_potentials}.
\end{align}
In other words, a suitably chosen set of transformations (\ref{eq:phaseTransformationGalilean}), (\ref{eq:scalarPotentialTransformationNonlinearGalileanTransformation}) and (\ref{eq:vectorPotentialTransformationNonlinearGalileanTransformation}) which solves $\xi=0$, signals the form invariance of the field equation
(\ref{eq:canonicalFieldEquationThetaDotSymmetry}).
Such a prospect is readily achieved by setting both Eqs. (\ref{eq:xi_phase_factor}) and (\ref{eq:xi_potentials}) to zero.
The condition, $\xi_{\chi}=0$, is upheld by a suitable phase transformation, while the imposition, $\xi_{\pi}=0$, leads to a transformation rule for the potentials.
Indeed, notice that the phase factor, $\chi=mw^2t/2-mw_iR_{ij}x_j$, solves the first of these two conditions, such that \cite{holland1995quantum,brown1999galilean}
\begin{equation}
    \theta\rightarrow\theta+\frac{1}{2}mw^2t-mw_iR_{ij}x_j,
  \label{eq:phase_transformation_galilean_elementary_fluid}
\end{equation}
while the second condition, is ensured for $\mu=-w_iR_{ij}A_j$.
Thus, the nonlinear potentials transform according to
\begin{align}
  \eta&\rightarrow\eta-w_iR_{ij}A_j, \label{eq:GalileanTransformationScalarPotentialNLGCF} \\
  A_i&\rightarrow R_{ij}A_j. \label{eq:GalileanTransformationVectorPotentialNLGCF}
\end{align}
These transformations are identical to those which appear for a Schr\"odinger field coupled to external scalar and vector potentials \cite{brown1999galilean}, but represent a nonlinear transformation. 
Note that the above transformations also appear for a classical charge, emerging as the magnetic limit of Galilean electromagnetism \cite{le1973galilean}.
In turn, performing transformations (\ref{eq:velocity_transformation_galilean}), (\ref{eq:GalileanTransformationScalarPotentialNLGCF}) and (\ref{eq:GalileanTransformationVectorPotentialNLGCF}) in Eq. (\ref{eq:nonlinearScalarPotentialWaveEquationTheta}) for the nonlinear scalar term in the wave equation for the phase, $\Phi$, yields
\begin{equation}
  \Phi\rightarrow\Phi-w_iR_{ij}A_j.
  \label{eq:GalileanTransformationQHJEscalarPotentialNLGCF}
\end{equation}
Hence $\Phi$ and $\eta$ undergo an identical transformation.
Furthermore, although the fluid pressure $P$ from Eq. (\ref{eq:pressureNLGCQF}) also depends explicitly on $\v{v}$, performing transformations (\ref{eq:GalileanTransformationScalarPotentialNLGCF}) and (\ref{eq:GalileanTransformationQHJEscalarPotentialNLGCF}) in Eq. (\ref{eq:pressureNLGCQF}) reveals that the covariant transformations of the potentials between frames ensure that the pressure is invariant.

\section{Conclusion}
The dynamical equations of a nonlinear gauge-coupled quantum fluid are invariant under local gauge transformations, insofar as the gauge function $\chi$ takes the form of an external scalar function of $\v{r}$ and $t$.
However, when $\chi$ becomes a functional of the density, the form-invariance of the dynamical equation for the phase $\theta$ is lost.
As a result, density-dependent gauge potentials are physical, non-trivial potentials which cannot be absorbed into the phase, since attempting to do so invariably destroys the form of the dynamical equations.
For a gauge-coupled one-dimensional superfluid, the additional term in the field equation generated by a nonlinear gauge transformation becomes integrable.
In particular, the gauge-pressure of the nonlinear fluid is generated in the Hamiltonian density for gauge transformations which eliminate the gauge potential from the kinetic energy density.
Finally, the immediate lack of Galilean covariance of the nonlinear fluid may be restored by subjecting the potentials to a nonlinear transformation.
As a result, the flow-dependent fluid pressure inherent to such systems, becomes invariant under a Galilean transformation.
These transformations are identical in form to those of a Schr\"odinger field subject to external scalar and vector potentials, but represent nonlinear transformations.

\begin{acknowledgements}
Y.B. acknowledges support from EPSRC CM-CDT Grant No. EP/L015110/1. P.\"O. acknowledges support from EPSRC EP/M024636/1. 
\end{acknowledgements}

\bibliographystyle{apsrev4-1}
%\bibliography{References2}

\end{document}